\documentclass[11pt]{article} 
\usepackage{graphicx,epsfig} 
\usepackage{geometry} 
\makeatletter 
\renewenvironment{thebibliography}[1] 
{\section*{\refname\@mkboth{\refname}{\refname}}%
  \list{\@biblabel{\@arabic\c@enumiv}}%
       {\settowidth\labelwidth{\@biblabel{#1}}%
        \leftmargin\labelwidth 
        \advance\leftmargin\labelsep 
 \setlength\baselineskip{11pt}%
        \@openbib@code 
        \usecounter{enumiv}%
        \let\p@enumiv\@empty 
        \renewcommand\theenumiv{\@arabic\c@enumiv}}%
  \sloppy 
  \clubpenalty4000 
  \@clubpenalty\clubpenalty 
  \widowpenalty4000%
  \sfcode`\.\@m} 
 {\def\@noitemerr 
 {\@latex@warning{Empty `thebibliography' environment}}%
\endlist} 
\makeatother 
\begin{document} 
\centerline{{\sl Genshikaku Kenkyu Suppl.} No. 000 (2012)}
\begin{center}  
\vskip 2mm 
{\Large\bf 
 
Present status of isobar models for elementary kaon photoproduction 
\hspace{-1mm}\footnote{Dedicated to our dear colleague and friend 
Miloslav Sotona who passed away April 6, 2012.}
\footnote{Presented at the International Workshop on Strangeness  
Nuclear Physics (SNP12), August 27 - 29, 2012, \\ 
\hspace*{5mm} Neyagawa, Osaka, Japan.} 
}\vspace{5mm} 
 
{ 
Petr Byd\v{z}ovsk\'y$^a$, Dalibor Skoupil$^a$ 
}\bigskip 
 
{\small 
$^a$Department of Theoretical Physics, Nuclear Physics Institute of the ASCR,  
\v{R}e\v{z}, 250 68, Czech Republic\\  
} 
\end{center} 
\vspace{3mm} 
 
\noindent 
{\small \textbf{Abstract}:\quad 
Some basic properties of isobar models are demonstrated and discussed 
on examples of the Saclay-Lyon and Kaon-MAID models. Predictions of these 
models are compared with experimental data for various processes 
and kinematical regions.
Results of the isobar models are also compared with the Regge-isobar 
hybrid model, Regge-plus-resonance.  
}%
 
 
\section{Introduction}
Photo- and electroproduction of kaons on nucleons and nuclei plays   
an important role in investigating the baryon spectra and  
interactions in the hyperon-nucleon systems. Studying details of 
dynamics of the strangeness electromagnetic production we can 
learn more about a nature and properties of nucleon resonances 
which supplements information obtained in other studies in 
the region of nonperturbative QCD. 
Proper understanding of reaction mechanism is also vital 
for the strangeness nuclear physics. Indeed, an accurate description 
of the elementary production amplitude is necessary, e.g. for reliable 
predictions of the cross sections in electroproduction 
of hypernuclei as the amplitude is part of an input which 
determines an accuracy of predictions~\cite{HProd}.

Description of the electroproduction process, 
$e + N \longrightarrow e' + K + Y$ ($N = n, p$; $K = K^+, K^0$; 
$Y = \Lambda, \Sigma^0, \Sigma^+, \Sigma^-$), 
can be formally reduced to investigation of a binary process 
of photoproduction by a virtual photon, 
$\gamma_V(q_\gamma) + N \longrightarrow K + Y$ 
($q_\gamma^2 < 0,\, Q^2 = -q_\gamma^2$),  
since the electromagnetic coupling constant is small enough  
to justify the one-photon exchange approximation. 
  
There are various ways how to describe the elementary 
electroproduction process. Among them the isobar model based 
on an effective Lagrangian description considering only hadronic 
degrees of freedom is suitable for more complex calculations of 
electroproduction of hypernuclei~\cite{HProd}. 
Another approach, suited also for description above the nucleon resonance 
region up to $\approx 16$ GeV, is the hybrid Regge-plus-resonance 
model~\cite{RPR,Lesley} (RPR). This model combines the Regge 
model~\cite{Regge}, appropriate to description above the resonance 
region ($E_{\gamma}>4$ GeV), with elements of the isobar model 
eligible for the low-energy region. 
In quark models for photoproduction~\cite{SagQM}, resonances are 
implicitly included as excited states and therefore a number of 
free parameters is relatively small.
Another asset of this approach is a natural description of a hadron 
structure which has to be modeled phenomenologically via form 
factors in the isobar models.
However, the quark models are too complicated for their further 
use in the calculations of hypernucleus electroproduction. 
 
In the effective hadrodynamical approach, various channels  
connected via the final-state meson-baryon interaction 
(rescattering processes) have to be treated simultaneously 
to take unitarity properly into account~\cite{CouCh}.   
In the coupled-channel approach, e.g. important effects 
of the $\pi$N intermediate states can be included.  
Considerable simplification originates in neglecting the 
rescattering effects in the formalism assuming that they  
are included to some extent  
by means of effective values of the coupling constants  
fitted to data. This simplifying assumption was adopted in many  
of the isobar models, e.g. Saclay-Lyon (SL)~\cite{SLA},  
Kaon-MAID (KM)~\cite{KM}, and Gent-Isobar~\cite{Jan01}.
 
\section{Isobar model}  
In the isobar model the amplitude is constructed by using 
the Feynman diagrammatic technique taking into account only 
contributions of the tree-level diagrams. The amplitude  
obtains contributions from the Born terms and $s$-,  
$t$-, and $u$-channel exchanges of the nucleon, kaon and hyperon 
resonances, respectively. 
Moreover, in some cases when gauge invariance is violated, e.g. 
due to hadronic form factors, a contact term is added to restore 
the gauge invariance. Absence of a dominant nucleon resonance 
in the kaon photoproduction (unlike in the $\pi$ and $\eta$ 
photoproduction) results in a large number of various combinations 
of the resonances with mass below 2 GeV~\cite{SLA} which fit 
the experimental data equally well.  
This plethora of models is limited considering constraints set by  
SU(3)~\cite{SLA,KM} and crossing symmetries~\cite{SLA}.
 
A drawback of the isobar model consists in a too large contribution 
of the Born terms to the non-resonant part of the amplitude 
(background)~\cite{Jan01}. To reduce this nonphysical contribution, 
either exchanges of the hyperon resonances are added~\cite{SLA} or 
the hadronic form factors (hff) in the strong vertexes are 
included~\cite{KM}. In the Gent-Isobar model a combination of 
both mechanisms is used~\cite{Jan01}.
Besides a reduction of the Born terms the hff can model an internal 
structure of hadrons in the strong vertexes which is neglected 
in the effective Lagrangian. 
The form factors are included by a gauge-invariant technique~\cite{DW} 
assuming dipole~\cite{KM,Jan01}, Gaussian~\cite{RPR} or 
multidipole-Gauss~\cite{Lesley} types. 
These various methods of reduction of the Born terms affect strongly 
the dynamics of isobar model. The problem of the large Born contribution 
is avoided in the RPR approach.   

The KM and Saclay-Lyon A (SLA)~\cite{SLA} models include the 
Born diagrams and contributions from exchanges of the $K^*(890)$ 
and $K_1(1270)$ resonances. The main coupling constants, 
$g_{KN\Lambda}$ and $g_{KN\Sigma}$, fulfill the limits of 
20\% broken SU(3) symmetry~\cite{SLA} in both models.  
These models differ in a choice of $s$- and $u$-channel resonances, 
in a treatment of the hadron structure, and in a set of experimental 
data to which the free parameters were adjusted.  
In the SLA model, one nucleon, $P_{13}(1720)$, and four hyperon 
resonances are included whereas in KM four nucleon, $S_{11}(1650)$, 
$P_{11}(1710)$, $P_{13}(1720)$, and $D_{13}(1895)$, and no hyperon 
resonances are assumed~\cite{KM}.  
In the SLA model hadrons are treated as point-like objects but  
in KM their internal structure is modeled by means of dipole-type hffs.  
The SLA and KM models provide reasonable results for 
photon energies below 2.2 GeV. 

\begin{figure} [htb!]  
\begin{center}  
\includegraphics[width=56mm,angle=270]{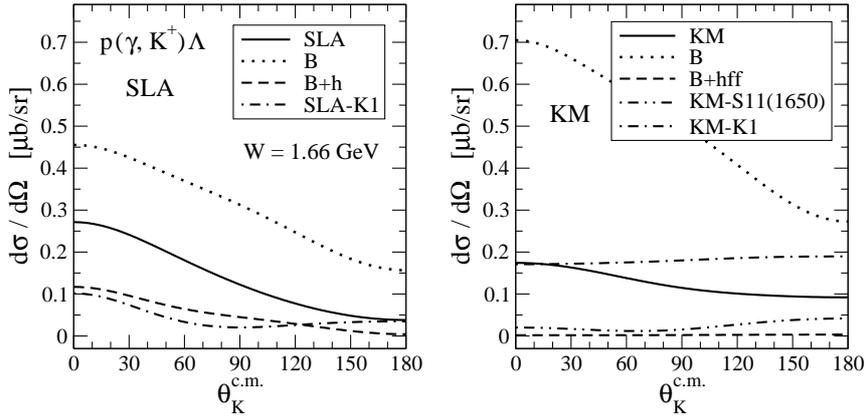}   
\end{center}  
\caption{\small Dynamics of the SLA (left) 
and KM (right) models in $p(\gamma,K^+)\Lambda$ at 
$E_\gamma^{lab}$ = 1~GeV. Contributions from the Born terms (B) 
are reduced by hyperon exchanges in SLA (B+h) 
and by the hff in KM (B+hff).
Solid and dash-doted lines show results of entire models and 
results without $K_1$ exchange, respectively.} 
\label{dynamics-KpL}  
\end{figure}

\begin{figure} [htb!]  
\begin{center}  
\includegraphics[width=56mm,angle=270]{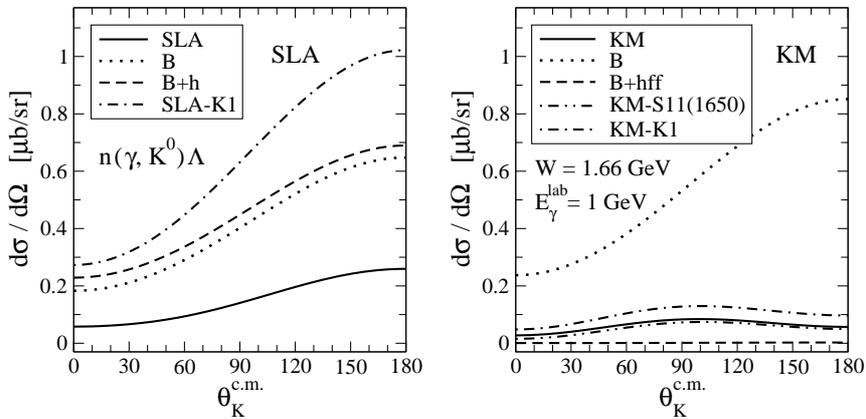}   
\end{center}  
\caption{\small The same as in Fig. 1 for $n(\gamma,K^0)\Lambda$. 
Value of the $K_1^0$ coupling constant in SLA is from Ref.~\cite{Tsukada}.} 
\label{dynamics-K0L}  
\end{figure}

Different dynamics of SLA and KM is shown in Figs.~\ref{dynamics-KpL} and 
\ref{dynamics-K0L} for $p(\gamma,K^+)\Lambda$ and $n(\gamma,K^0)\Lambda$, 
respectively. 
In Fig.~\ref{dynamics-KpL} the large forward-angle-peaked contribution  
from the Born terms without hff (B) is counterbalanced by the hyperon 
exchanges in SLA (B+h, left panel) and suppressed by the hff in 
KM (B+hff, right panel). 
In the KM model the contribution from the Born terms with hff is 
negligibly small which results in a dominance of contributions from 
the nucleon exchanges, see dash-double-doted line in the right panel 
for the entire KM without $S_{11}(1650)$. 
A significant difference in dynamics stems from the $K_1$ exchange. 
Results without the $K_1$ exchange (dash-doted lines) deviate from 
the full results at forward and backward angles for SLA and KM, 
respectively. 
%
%

In $n(\gamma,K^0)\Lambda$ the Born terms reveal 
different angular dependence due to absence of the kaon exchange and 
the electric part of the proton diagram. Moreover, the anomalous 
magnetic moment has an opposite sign: 
$\mu_p = 1.79 \rightarrow \mu_n=-1.92$. This results in the 
backward peaked contribution of the Born terms, 
Fig.~\ref{dynamics-K0L}. The contribution from hyperon resonances 
in SLA does not change the angular dependence (left panel) whereas 
the hff in KM reduces significantly the Born terms  
as in $p(\gamma,K^+)\Lambda$ (right panel). 
This causes the completely different angular dependence predicted by 
SLA and KM for $n(\gamma,K^0)\Lambda$. 
A role of the $K_1^0$ exchange is also very different in these models.  
In SLA the $K_1^0$ exchange counterbalances the Born terms, which 
makes SLA very sensitive to a value of the $K_1^0$ coupling constant, 
whereas in KM the $K_1^0$ exchange gives a very small contribution. 
In Fig.~\ref{dynamics-K0L} results of entire models without the $K_1^0$ 
exchange are shown as dash-doted lines. 
%
%
 
\section{Regge-plus-resonance model} 
\begin{figure} [htb!]  
\begin{center}  
\includegraphics[width=57mm,angle=270]{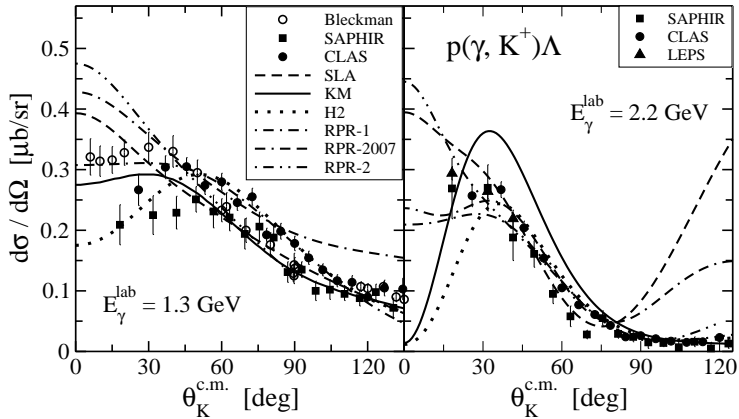}   
\end{center}  
\caption{\small Comparison of isobar and RPR models with data for 
$p(\gamma,K^+)\Lambda$ and photon energies 1.3 and 2.2~GeV. 
Data are from Refs.~\cite{Bleck}(Bleckman), \cite{SAPHIR03}(SAPHIR), 
\cite{CLAS05}(CLAS), \cite{LEPS}(LEPS).} 
\label{comp_isobar-RPR}  
\end{figure}

In the RPR model the non-resonant part of the amplitude is modeled by 
exchanges of two degenerate $K$ and $K^*$ trajectories. The three free 
parameters can be fixated from fitting to photoproduction data above 
the resonance region~\cite{RPR, Lesley}. 
The resonant part is described by exchanges of nucleon resonances like 
in the isobar model. A smooth transition from the resonant region into 
the high-energy Regge region is assured by strong hffs of Gaussian or 
multidipole-Gauss type~\cite{RPR,Lesley}. 
Important merit of the RPR model, besides that it describes data in 
the energy region up to $E_\gamma^{lab}\approx 16$ GeV, is absence of large Born contributions in the non-resonant part of the amplitude. 
Therefore, no hffs for the background 
are needed which makes a difference between the RPR model and isobar 
model with hff, which is important for very small kaon angles, see 
Fig.~\ref{comp_isobar-RPR}.

In Fig.~\ref{comp_isobar-RPR} predictions for photoproduction of $K^+$ 
are compared for isobar models with (KM, H2~\cite{HYP03}) and without (SLA) 
hff, and for three versions of the RPR model, RPR-2007~\cite{RPR} and 
our recent versions RPR-1 and RPR-2, motivated by RPR-2011B~\cite{Lesley} and 
fitted to CLAS and LEPS data below 2.5 GeV. The largest differences appear for 
the energy 2.2 GeV at $\theta_K^{c.m.} < 30^\circ$. The isobar models 
with hff (KM, H2) predict very small cross sections at zero angle 
and a steeply rising angular dependence, which is given by a strong 
suppression of the proton exchange by hff, whereas the models without 
hff (SLA) give a decreasing dependence and large cross sections. 
The RPR models can give either a plato (RPR-2007, RPR-1) or a steeply 
decreasing dependence (RPR-2) similarly as the SLA model . 
%
%

\section{Results} 
      
\begin{figure} [t!]   
\begin{center}     
\includegraphics[width=47mm,angle=270]{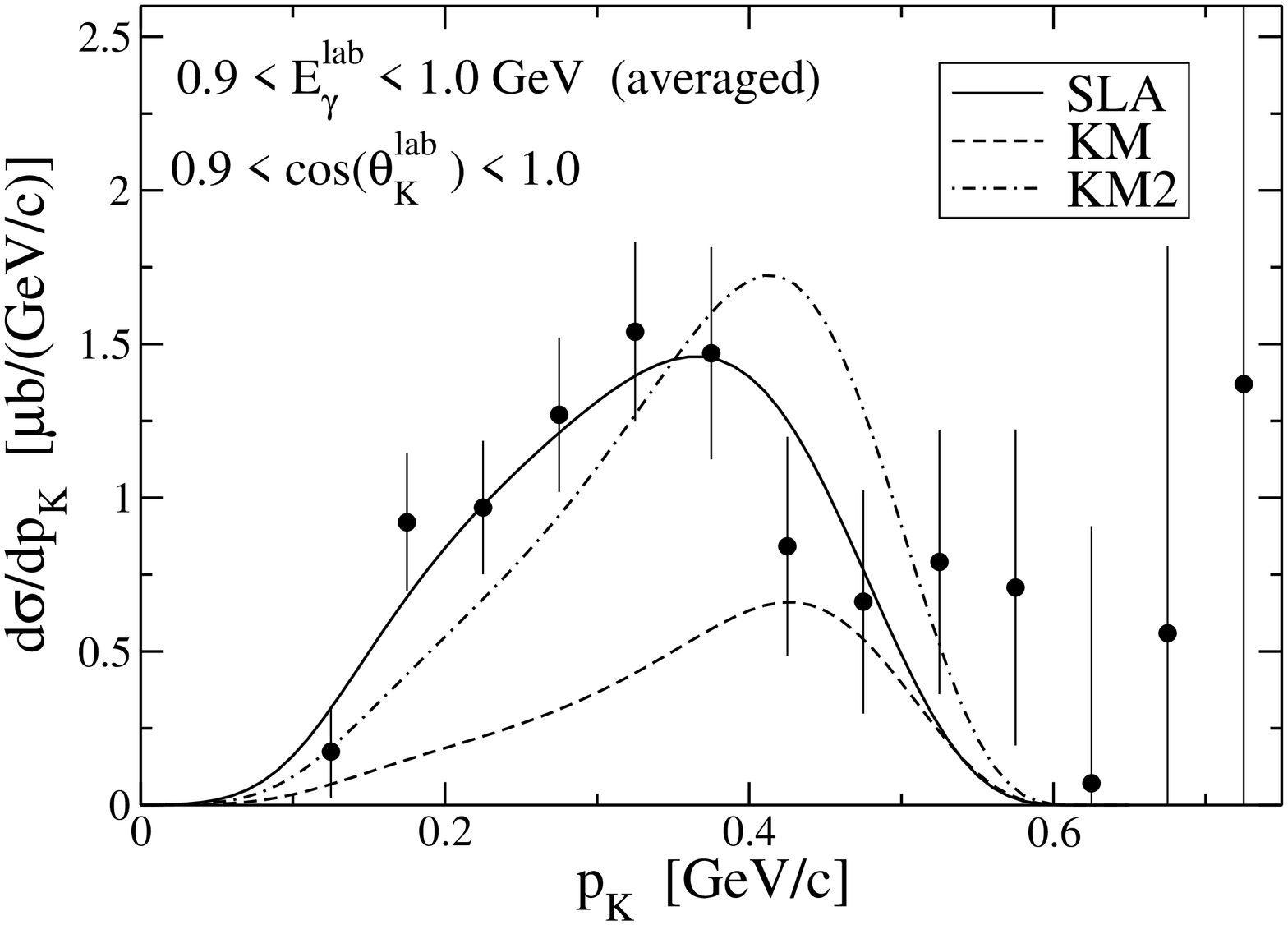}
\hspace{3mm}
\includegraphics[width=47mm,angle=270]{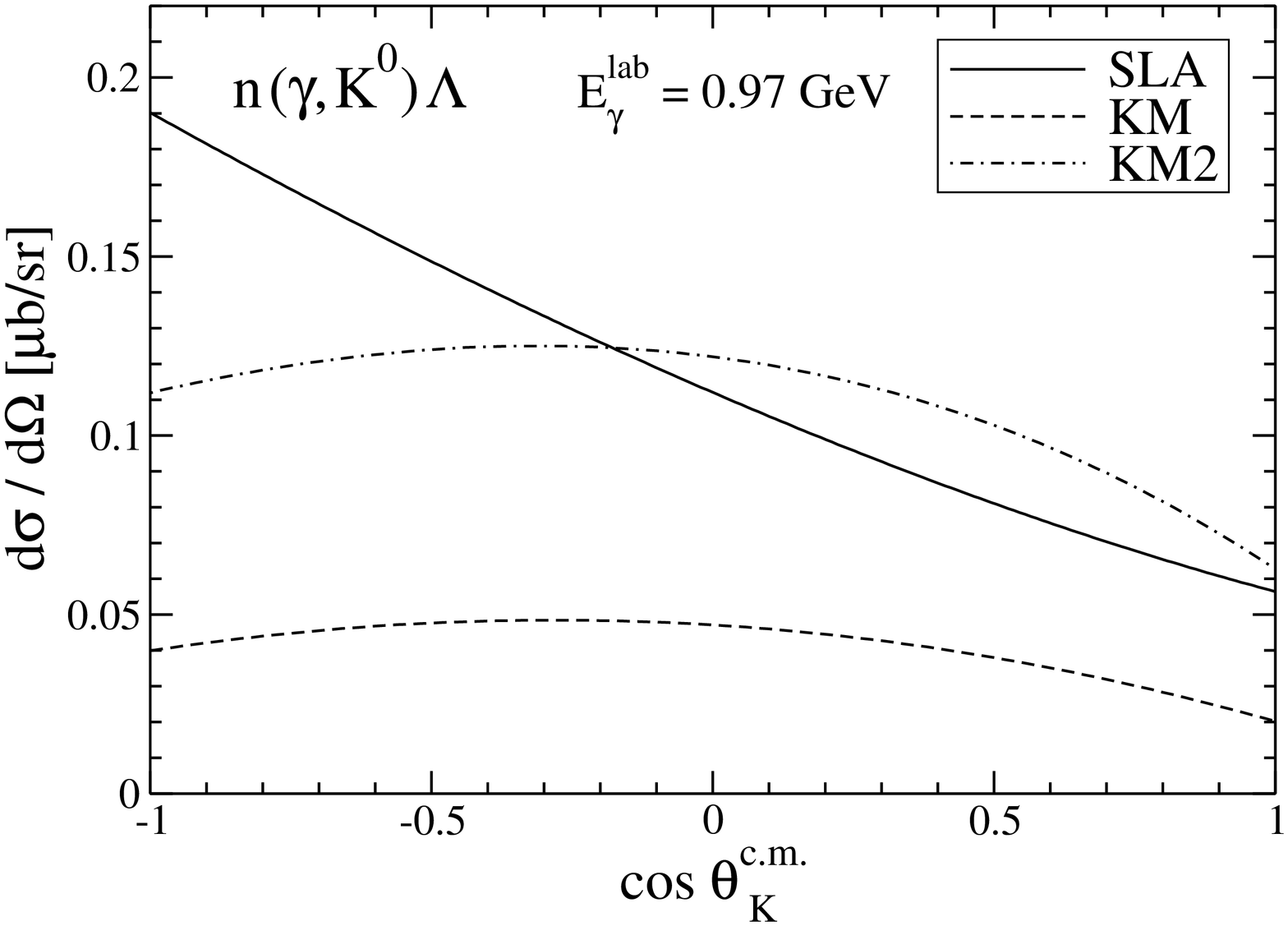}   
\end{center}   
\caption{\small PWIA calculations of the energy-averaged 
and kaon-angle integrated momentum distribution for 
$d(\gamma,K^0$)$\Lambda p$ 
are compared with data~\cite{Tsukada} (left). Angular dependence of the 
elementary cross sections in c.m.s. at the central energy is shown for 
the SLA~\cite{Tsukada}, 
KM~\cite{KM} models and fitted version KM2 with $r_{K_1K\gamma}$ = 0.47.} 
\label{K0-deuteron}  
\end{figure}

The differences in dynamics of the SLA and KM models in the 
$n(\gamma,K^0)\Lambda$ reaction, shown in Fig.~\ref{dynamics-K0L}, 
can be checked comparing the PWIA calculations with data for 
the photoproduction of $K^0$ on deuteron~\cite{Tsukada}. 
In Fig.~\ref{K0-deuteron} results of SLA and two versions of Kaon-MAID 
model, KM and KM2, for the energy-averaged and angle-integrated momentum 
distribution of the $K^0\Lambda$ production are shown for 
the energy region below 1 GeV where contributions from 
the $\Sigma$ production are negligible. Results of SLA and KM2 were fitted 
to the data optimizing the $K_1^0$ coupling constant, see Ref.~\cite{Tsukada} 
for SLA, whereas KM was calculated 
with the original coupling constant fitted to $K^0\Sigma$ data~\cite{KM}. 
It is apparent that SLA fits the shape of distribution very well 
($\chi^2/n.d.f.$ = 0.64). The Kaon-MAID model can fit only a magnitude of 
data points but cannot change the shape of distribution 
($\chi^2/n.d.f.$ = 1.75 for KM2). 
%
%
Angular dependence of the elementary cross sections is shown on the right panel 
of Fig.~\ref{K0-deuteron}. Both fitted models, SLA and KM2, give a similar 
magnitude of the cross section but very different angular dependence which is 
given by their dynamics.
The shape of angular dependence is important for a good description of 
the momentum distribution in the 
$d(\gamma,K^0)\Lambda p$ reaction, see also Ref.~\cite{Tsukada}.

\begin{figure} [htb!]  
\begin{center}  
\includegraphics[width=46mm,angle=270]{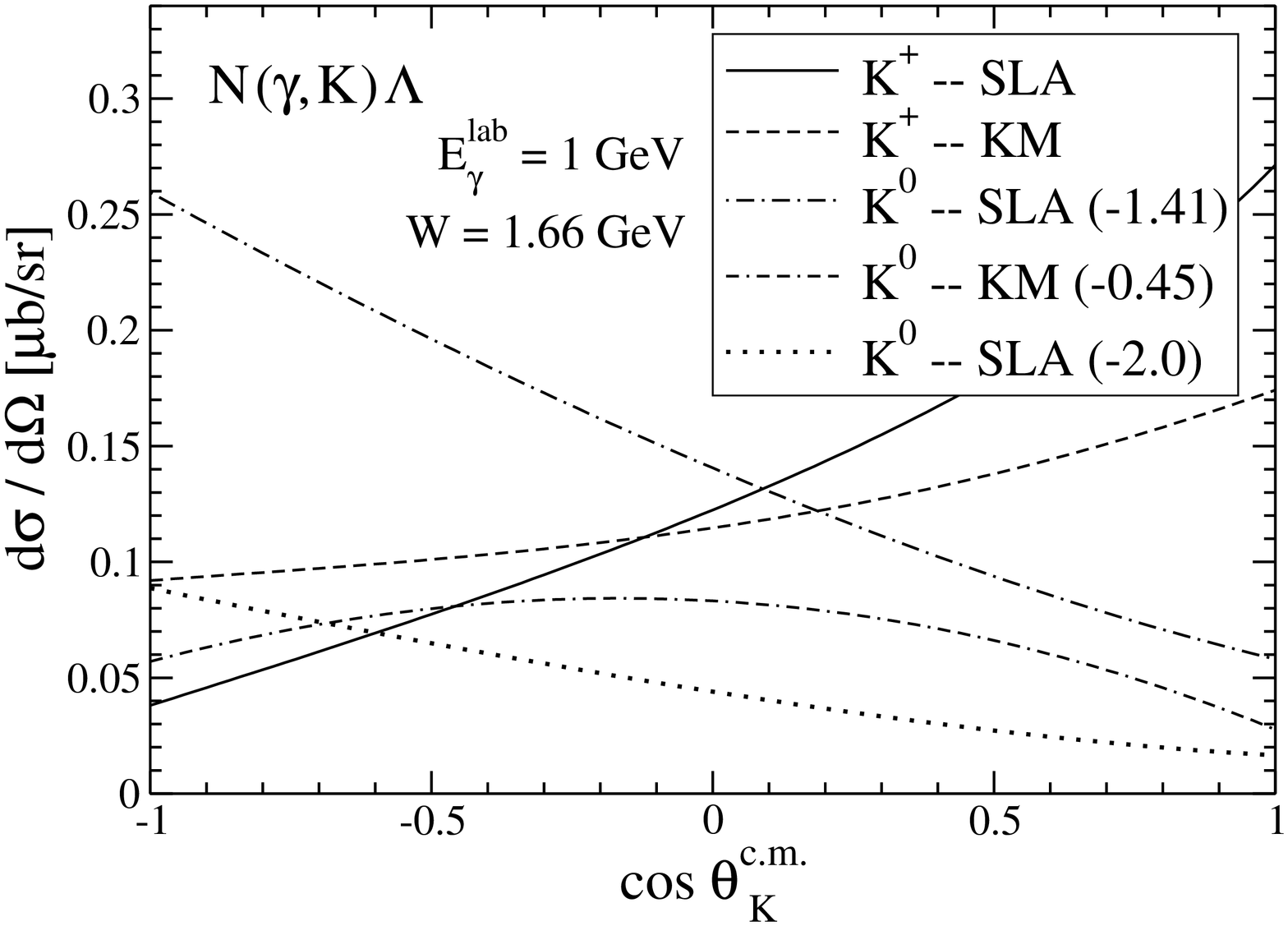} 
\hspace{3mm} 
\includegraphics[width=47mm,angle=270]{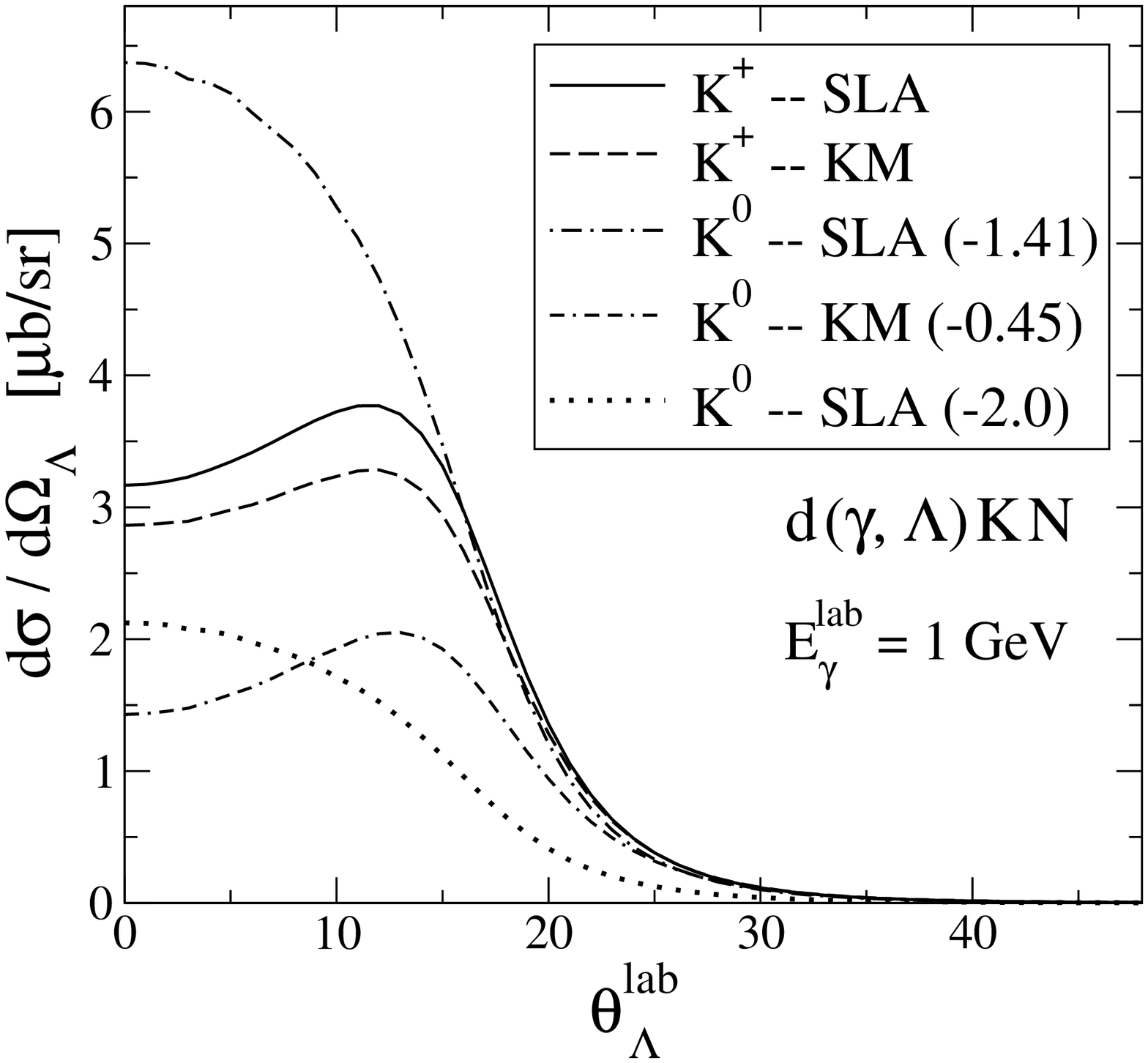}   
\end{center}  
\caption{\small Cross sections for $p(\gamma,K^+)\Lambda$ 
and $n(\gamma,K^0)\Lambda$ predicted by the KM and two versions 
of SLA models (left). Values of $r_{K_1K\gamma}$ are in parenthesis. 
Predictions of the $\Lambda$-momentum-integrated cross sections 
for $d(\gamma,\Lambda)K^+n$ and $d(\gamma,\Lambda)K^0p$ 
as a function of $\Lambda$ laboratory angle using the elementary  
amplitudes (right).} 
\label{L-deuteron}  
\end{figure}

In Fig.~\ref{L-deuteron} predictions for photoproduction on nucleons 
(left panel) and the deuteron (right panel) with the $K^+\Lambda$ 
and $K^0\Lambda$ final states are shown for KM and two variants 
of SLA models, which differ in values of the $r_{K_1K\gamma}$ 
parameter~\cite{Tsukada} shown in parenthesis in the figure.
The SLA model appears to be very sensitive to the parameter 
$r_{K_1K\gamma}$ which can modify both magnitude and angular 
dependence of the cross section. 
This sensitivity was also observed in Fig.~\ref{dynamics-K0L}. 
The SLA with $r_{K_1K\gamma}$ = -1.41 predicts 
the $\Lambda$-momentum-integrated cross section to be dominated by 
the $K^0$ production.
%
%
 
\begin{figure} [t!]  
\begin{center}  
\includegraphics[width=50mm,angle=270]{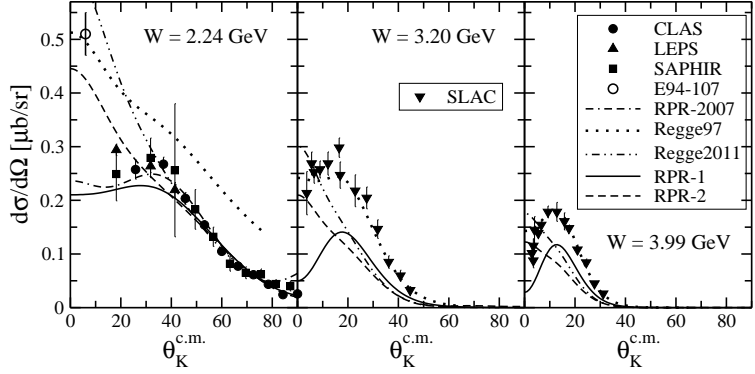}   
\end{center}  
\caption{\small Results of Regge and Regge-plus-resonance models 
for the cross sections in $p(\gamma,K^+)\Lambda$ are compared 
with CLAS~\cite{CLAS05} and SLAC~\cite{SLAC} data at forward-angle 
region and for energies above 2.2 GeV. The Jlab Hall A data point 
(E94-107)~\cite{E89-009} is for electroproduction very near 
to the photoproduction point, $Q^2$ = 0.07~(GeV/c)$^2$.} 
\label{small_ang}  
\end{figure}

In Fig.~\ref{small_ang} results of the Regge and RPR models for the 
forward kaon angles are compared with data for energies above W = 2.2 GeV. 
The problem of normalization of SLAC data~\cite{SLAC_norm} is apparent 
from the energy dependence of Regge97 fitted to these data~\cite{Regge}. 
The new version, Regge2011 fitted by Gent group to CLAS data above 
2.6 GeV~\cite{Lesley}, is consistent with low-energy data but 
underpredicts SLAC data. 
The SLAC data show a suppression of the cross section at zero angle  
and W = 3.99 GeV which is not observed in the low-energy region 
(W = 2.24 GeV). On the contrary, the electroproduction data point,  
E94-107, which is near to photoproduction, 
$Q^2$~=~0.07~(GeV/c)$^2$~\cite{E89-009}, suggests rather a decreasing 
angular dependence. 
This behavior is predicted by Regge2011 and RPR-2 models. 
The RPR-1 and RPR-2007~\cite{RPR} models suggest rather a plato at very small 
angles.
%
%
     
%
%
\begin{figure} [b!]  
\begin{center}  
\includegraphics[width=51mm,angle=270]{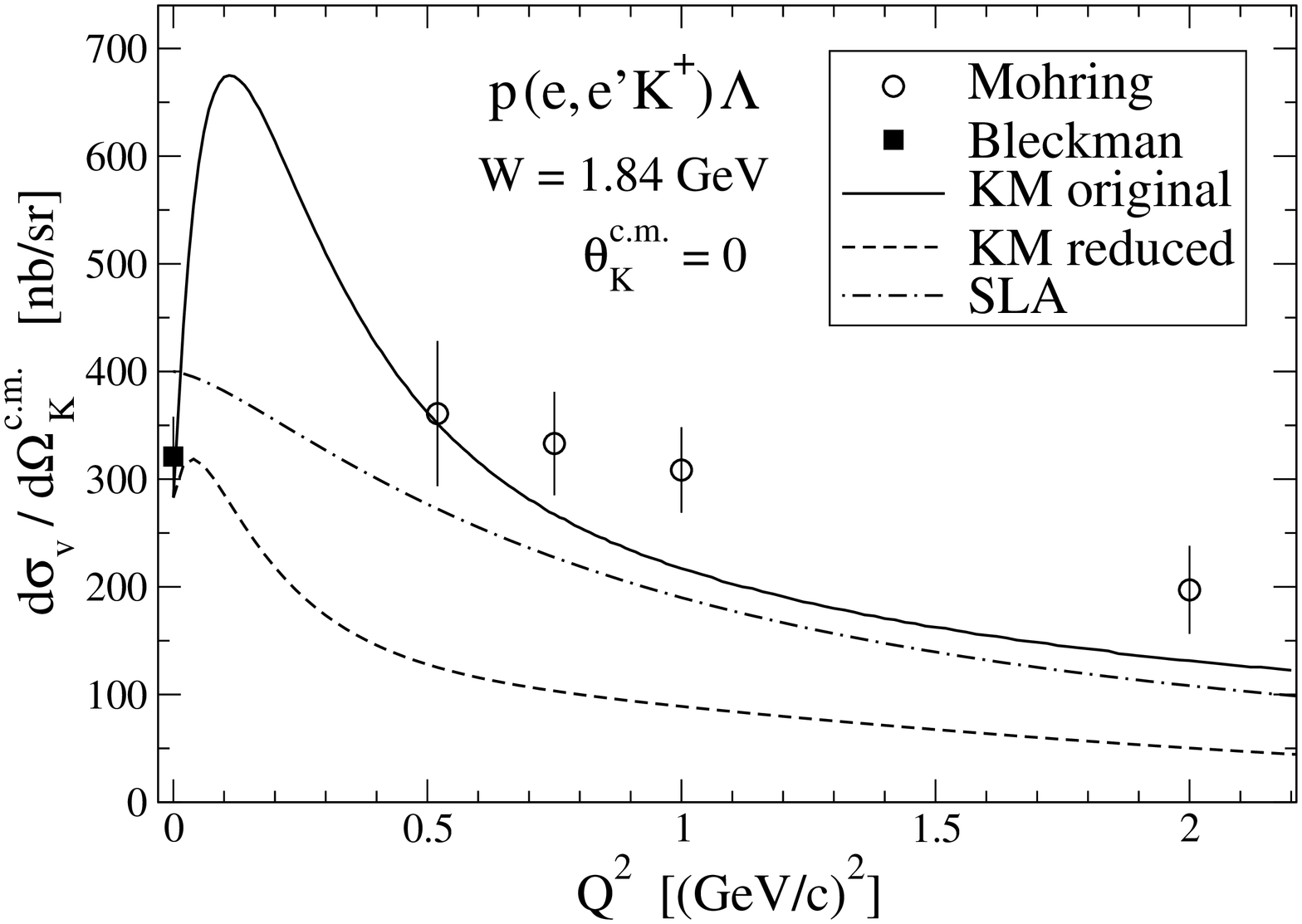}
\hspace{3mm}  
\includegraphics[width=51mm,angle=270]{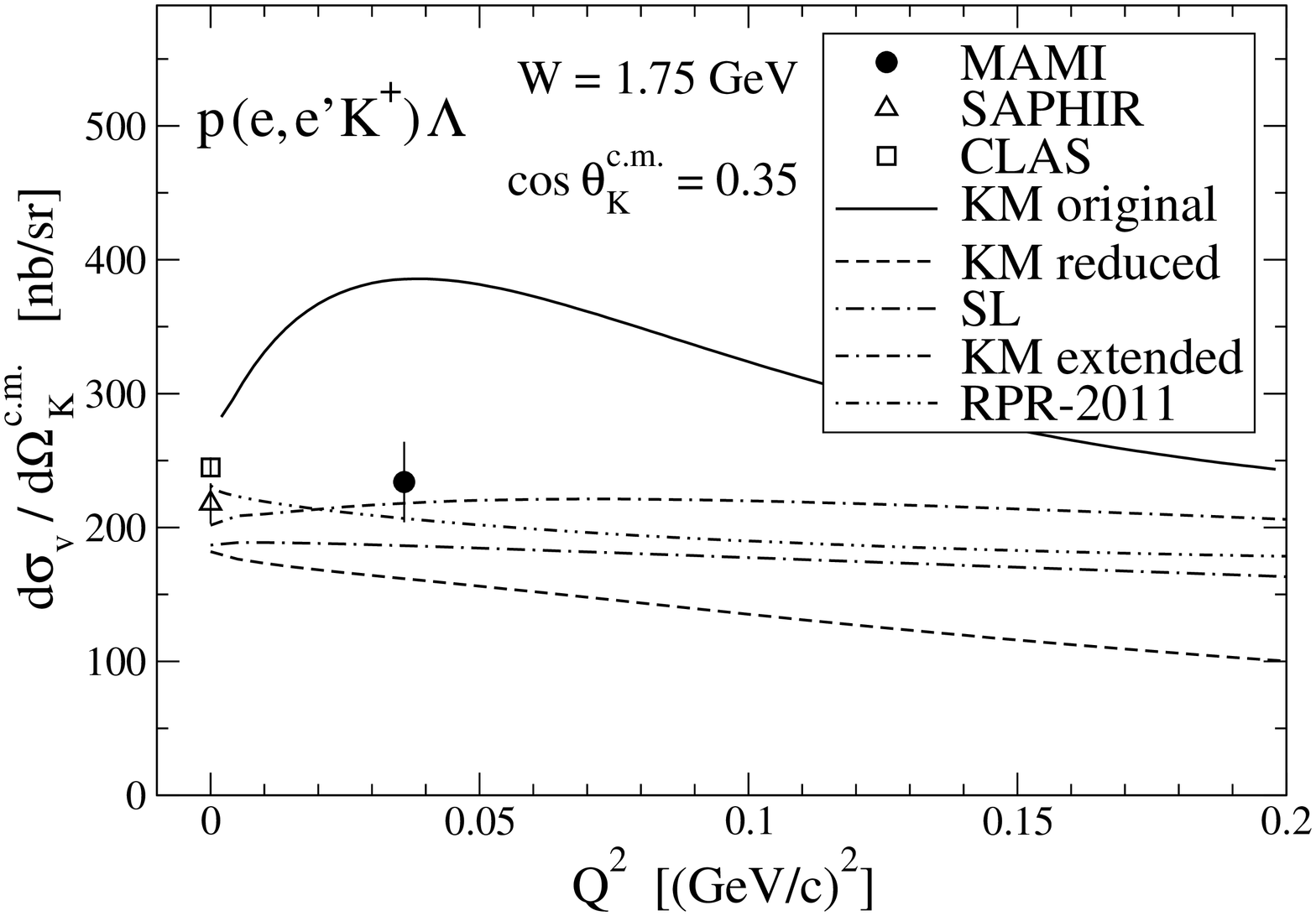}   
\end{center}  
\caption{\small Predictions of the isobar models for the electroproduction 
cross section are compared with data to show behavior near the photoproduction 
point ($Q^2$=0).} 
\label{small_Q2}  
\end{figure}      
In electroproduction one should include in the effective Lagrangian also 
other possible couplings of virtual photon with baryons, e.g. 
``longitudinal couplings" (LC). 
The corresponding coupling constants can be established by fitting 
the $Q^2$ dependence of the electroproduction cross sections. 
This was done for the KM model using data by Niculescu 
et al~\cite{JLab}. The KM result, ``KM original", 
for the full unpolarized cross section with 
$\epsilon$ = 0.5 is shown in Fig.~\ref{small_Q2} (left) in 
comparison with the SLA model and the re-analyzed data by Mohring 
et al~\cite{JLab}. The sharp bump for 0 $< Q^2 <$ 0.5 (GeV/c)$^2$ 
is modeled in KM by strong LC as it is apparent from 
a comparison with the result of ``KM reduced" in which these 
couplings were removed. The SLA model which does not include LC 
predicts a smooth $Q^2$ dependence. 
Behavior of the cross section near the photoproduction 
point ($Q^2$ = 0) is shown on the right panel of  Fig.~\ref{small_Q2}. 
The MAMI data~\cite{MAMI} at energy W = 1.75 GeV suggest a smooth   
$Q^2$ dependence contrary to that predicted by the original KM. 
The new version 
``KM extended" with reduced values of LC is very well consistent 
with the new data and the RPR-2011~\cite{Lesley} and SL models, see also 
Ref.~\cite{MAMI}.   
  
\section{Summary} 
Data on the $K^0$ photoproduction on deuteron can be used for testing 
the isobar models. More data for very small kaon angles and wide energy range  
are needed to shed light on the angular and energy dependence of the cross 
section and dynamics of isobar models in this kinematical region. 
Recent data for very small $Q^2$ suggest that longitudinal couplings 
are not too much significant in the isobar models.
  
\noindent ACKNOWLEDGMENT: 
P.B. thanks the organizers for their invitation to the conference. 
This report was supported by the Grant Agency of  
the Czech Republic, grant P203/12/2126.

\end{document}